\documentclass[aps,prl,
reprint,
showpacs,
amsmath,amssymb,
superscriptaddress,
altaffilletter
]{revtex4-1}
\usepackage{graphicx}
\begin{document}


\title{Precise determination of $^{12}_{\Lambda}$C level structure by $\gamma$-ray spectroscopy}


\author{K.~Hosomi}
\altaffiliation[Present Address: ]{Advanced Science Research Center (ASRC), Japan Atomic Energy Agency (JAEA), Ibaraki 319-1195, Japan}
\affiliation{Department of Physics, Tohoku University, Sendai 980-8578, Japan}

\author{Y.~Ma}
\altaffiliation[Present Address: ]{Advanced Meson Science Laboratory, RIKEN, Wako 351-0198, Japan}
\affiliation{Department of Physics, Tohoku University, Sendai 980-8578, Japan}

\author{S.~Ajimura}
\altaffiliation[Present Address: ]{Research Center for Nuclear Physics (RCNP), Osaka University, Ibaraki 567-0047, Japan}
\affiliation{Department of Physics, Osaka University, Toyonaka 560-0043, Japan}

\author{K.~Aoki}
\affiliation{High Energy Accelerator Research Organization (KEK), Tsukuba 305-0801, Japan}

\author{S.~Dairaku}
\affiliation{Department of Physics, Kyoto University, Kyoto 606-8502, Japan}

\author{Y.~Y.~Fu}
\affiliation{China Institute of Atomic Energy, Beijing 102413, China}

\author{H.~Fujioka}
\altaffiliation[Present Address: ]{Department of Physics, Kyoto University, Kyoto 606-8502, Japan}
\affiliation{Department of Physics, University of Tokyo, Tokyo 113-0033, Japan}

\author{K.~Futatsukawa}
\altaffiliation[Present Address: ]{High Energy Accelerator Research Organization (KEK), Tsukuba 305-0801, Japan}
\affiliation{Department of Physics, Tohoku University, Sendai 980-8578, Japan}

\author{W.~Imoto}
\affiliation{Osaka Electro-Communication University, Neyagawa 572-8530, Japan}

\author{Y.~Kakiguchi}
\affiliation{High Energy Accelerator Research Organization (KEK), Tsukuba 305-0801, Japan}

\author{M.~Kawai}
\affiliation{Department of Physics, Tohoku University, Sendai 980-8578, Japan}

\author{S.~Kinoshita}
\affiliation{Department of Physics, Tohoku University, Sendai 980-8578, Japan}

\author{T.~Koike}
\affiliation{Department of Physics, Tohoku University, Sendai 980-8578, Japan}

\author{N.~Maruyama}
\affiliation{Department of Physics, Tohoku University, Sendai 980-8578, Japan}

\author{M.~Mimori}
\affiliation{Department of Physics, Tohoku University, Sendai 980-8578, Japan}

\author{S. Minami}
\altaffiliation[Present Address: ]{GSI Helmholtz Center for Heavy Ion Research, Darmstadt, Germany}
\affiliation{Department of Physics, Osaka University, Toyonaka 560-0043, Japan}

\author{Y.~Miura}
\affiliation{Department of Physics, Tohoku University, Sendai 980-8578, Japan}

\author{K.~Miwa}
\altaffiliation[Present Address: ]{Department of Physics, Tohoku University, Sendai 980-8578, Japan}
\affiliation{Department of Physics, Kyoto University, Kyoto 606-8502, Japan}

\author{Y.~Miyagi}
\affiliation{Department of Physics, Tohoku University, Sendai 980-8578, Japan}

\author{T.~Nagae}
\altaffiliation[Present Address: ]{Department of Physics, Kyoto University, Kyoto 606-8502, Japan}
\affiliation{High Energy Accelerator Research Organization (KEK), Tsukuba 305-0801, Japan}

\author{D.~Nakajima}
\altaffiliation[Present Address: ]{Institute for Cosmic Ray Research, University of Tokyo, Kashiwa 227-8582, Japan}
\affiliation{Department of Physics, Kyoto University, Kyoto 606-8502, Japan}

\author{H.~Noumi}
\altaffiliation[Present Address: ]{Research Center for Nuclear Physics (RCNP), Osaka University, Ibaraki 567-0047, Japan}
\affiliation{High Energy Accelerator Research Organization (KEK), Tsukuba 305-0801, Japan}

\author{K.~Shirotori}
\altaffiliation[Present Address: ]{Research Center for Nuclear Physics (RCNP), Osaka University, Ibaraki 567-0047, Japan}
\affiliation{Department of Physics, Tohoku University, Sendai 980-8578, Japan}

\author{T.~Suzuki}
\altaffiliation[Present Address: ]{Research Center for Nuclear Physics (RCNP), Osaka University, Ibaraki 567-0047, Japan}
\affiliation{Cyclotron and Radioisotope Center, Tohoku University, Sendai 980-8578, Japan}

\author{T.~Takahashi}
\affiliation{High Energy Accelerator Research Organization (KEK), Tsukuba 305-0801, Japan}

\author{T.~N.~Takahashi}
\altaffiliation[Present Address: ]{Research Center for Nuclear Physics (RCNP), Osaka University, Ibaraki 567-0047, Japan}
\affiliation{Department of Physics, University of Tokyo, Tokyo 113-0033, Japan}

\author{H.~Tamura}
\affiliation{Department of Physics, Tohoku University, Sendai 980-8578, Japan}

\author{K.~Tanida}
\altaffiliation[Present Address: ]{Department of Physics and Astronomy, Seoul National University, Seoul 151-747, Republic of Korea}
\affiliation{RIKEN, Wako 351-0198, Japan}

\author{N.~Terada}
\affiliation{Department of Physics, Tohoku University, Sendai 980-8578, Japan}

\author{A.~Toyoda}
\affiliation{High Energy Accelerator Research Organization (KEK), Tsukuba 305-0801, Japan}

\author{K.~Tsukada}
\altaffiliation[Present Address: ]{Research Center for Electron Photon Science (ELPH), Tohoku University, Sendai 982-0826, Japan}
\affiliation{Department of Physics, Tohoku University, Sendai 980-8578, Japan}

\author{M.~Ukai}
\altaffiliation[Present Address: ]{Department of Physics, Tohoku University, Sendai 980-8578, Japan}
\affiliation{Cyclotron and Radioisotope Center, Tohoku University, Sendai 980-8578, Japan}

\author{S.~H.~Zhou}
\affiliation{China Institute of Atomic Energy, Beijing 102413, China}

\collaboration{Hyperball (KEK-E566) Collaboration}
\noaffiliation

\date{\today}

\begin{abstract}
Level structure of a $^{12}_{\Lambda}$C hypernucleus was precisely determined by means of $\gamma$-ray spectroscopy.
We identified four $\gamma$-ray transitions via the $^{12}$C$(\pi^{+},K^{+}\gamma)$ reaction using a germanium detector array, Hyperball2.
The spacing of the ground-state doublet $(2_{1}^{-},1^{-}_{1})$ was measured to be $161.5\pm0.3\text{(stat)}\pm0.3\text{(syst)}$\,keV from the direct $M1$ transition.
Excitation energies of the $1^{-}_{2}$ and $1^{-}_{3}$ states were measured to be $2832\pm3\pm4$\,keV and $6050\pm8\pm7$\,keV, respectively.
The obtained level energies provide a definitive reference for the reaction spectroscopy of $\Lambda$ hypernuclei.

\end{abstract}


\maketitle




%

Spectroscopic studies of hypernuclei have played significant roles in our understanding of hyperon-nucleon (YN) and hyperon-hyperon (YY) interactions, since short lifetimes of hyperons make YN and YY scattering experiments technically difficult unlike in the nucleon-nucleon (NN) case.

Since 1970's structures of $\Lambda$ hypernuclei have been investigated through various types of reaction spectroscopy, in which hypernuclear states are directly populated and analyzed by the $(K^{-},\pi^{-})$, $(\pi^{+},K^{+})$, and $(e,e'K^{+})$ reactions.
The first reaction spectroscopy experiment utilized the $(K^{-}_{\text{stopped}},\pi^{-})$ reaction at CERN~\cite{faessler_1973_pl}, which was soon followed by the in-flight $(K^{-},\pi^{-})$ reaction~\cite{bruckner_1975_plb} with a small momentum transfer. 
After 1980's the $(\pi^{+},K^{+})$ reaction with a large momentum transfer was developed at BNL~\cite{milner_1985_prl} and later at KEK~\cite{akei_1991_npa}.
At the beginning of this century, the $(e,e'K^{+})$ reaction has become possible at Jefferson Lab (JLab)~\cite{miyoshi_2003_prl}, achieving a sub-MeV (FWHM) mass resolution.
In addition, a new type of experiments was also performed with the $(K^{-}_{\text{stopped}},\pi^{-})$ reaction with an excellent resolution using low-energy $K^-$s from $\phi$ meson decay at the DA$\Phi$NE facility~\cite{agnello_2005_plb}.
Those reaction methods have their own characteristics in terms of selectivity of populating states, mass resolution, experimental yields, background, and so on.
Among them one can choose the most suitable reaction method according to the purpose of the study.
Although it is necessary to cross check and/or to calibrate the mass scale among different experiments using those different reaction methods, in general it is not straightforward because populated states are not necessarily the same between the different reactions.

\begin{figure}
\includegraphics[width=6cm]{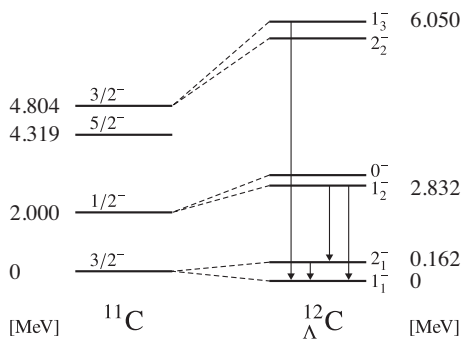}
\caption{\label{level}Low-lying level structures of $^{12}_{\Lambda}$C with the corresponding levels for the core nucleus, $^{11}$C~\cite{toi_npa}. The four $\gamma$-ray transitions observed in the present experiment (KEK-PS E566) are shown together with the three excitation energies determined.  }
\end{figure}

In all of those reaction spectroscopy experiments, carbon ($^{12}$C) has been chosen as the first target and then used to obtain a benchmark spectrum of $^{12}_\Lambda$C or $^{12}_\Lambda$B for verification of the spectrometer performance.
This is because 1) its structure is rather simple, and two peaks corresponding to hypernuclear states with a $\Lambda$ in the $s$ and the $p$ orbit coupled to the core $^{11}$C ground state $(3/2^{-})$ can be clearly separated, 2) the two peaks have relatively large production cross sections, and 3) $^{12}$C is readily available as a target.
The $(\pi^{+},K^{+})$ experiments at KEK-PS~\cite{hasegawa_1995_prl,hotch_2001_prc} revealed small peaks lying between the prominent two peaks in the $^{12}_{\Lambda}$C spectrum for the first time.
These were assigned as the core excited states where a $\Lambda$ in the $s$ orbit is coupled to the excited states $(1/2^-, 3/2^-)$ of $^{11}$C (Fig.~\ref{level}).
Similar peaks were also observed later in $^{12}_\Lambda$B spectra via the $(e,e'K^+)$ reaction~\cite{iodice_2007_prl,tang_2014_prc}. 
\begin{table*}
\caption{\label{state_table} List of the low-lying excitation energies of $^{12}_{\Lambda}$C ($^{12}_{\Lambda}$B) measured in various experiments. The spin-parities of the states which are expected to be populated theoretically are given in the $J^{\pi}$ column based on the weak-coupling picture of a $\Lambda$ and a $^{11}$C($^{11}$B) core. ``Present'' shows the results obtained from $\gamma$-ray spectroscopy (KEK-PS E566), and the others are measured by missing mass spectroscopy.}
\begin{ruledtabular}
\begin{tabular}{cccccccc}
  & Present & KEK-PS E369~\cite{hotch_2001_prc} & FINUDA~\cite{agnello_2005_plb} & \phantom{AA} & &JLab Hall A~\cite{iodice_2007_prl} & JLab Hall C~\cite{tang_2014_prc} \\
  & $(\pi^{+}, K^{+}\gamma)$ & $(\pi^{+}, K^{+})$ & $(K^{-}_{\text{stopped}}, \pi^{-})$ & & & \multicolumn{2}{c}{$(e,e'K^{+})$} \\
  $J^{\pi}$ & \multicolumn{3}{c}{$^{12}_{\Lambda}$C Excitation Energy (MeV)} & & $J^{\pi}$ &  \multicolumn{2}{c}{$^{12}_{\Lambda}$B Excitation Energy (MeV)}\\
 \hline
 $1^{-}_{1}$ & 0.0  & 0.0 & $0.0\pm0.06$ & &  $1^{-}_{1}$ & $\phantom{0}0.0\pm0.03$ & $\phantom{000}0.0\pm0.019$\\
 $2^{-}_{1}$ & $0.1615\pm0.0003$  & - & -\phantom{0}  &&  $2^{-}_{1}$ & \phantom{$^{a}$}0.65\footnotemark[1] & \phantom{$0^{a}$}$0.179\pm0.027$\footnote{Peaks for $1^{-}_{1}$ and $2^{-}_{1}$ are not resolved but their energy spacing is derived via a fitting by assuming two peaks with some constrains. See references for details.} \\
 $1^{-}_{2}$ & $2.832\pm0.003$ & $2.51\pm0.17$ & $2.5\pm0.2\phantom{0}$  & & $1^{-}_{2}$,$0^{-}$ & $2.65\pm0.10$ & $\phantom{0}3.109\pm0.044$ \\
 $1^{-}_{3}$ & $6.050\pm0.010$  & $6.30\pm0.11$ & $5.0\pm0.1\phantom{0}$  & &  $2^{-}_{2}$,$1^{-}_{3}$ & $5.92\pm0.13$ & $\phantom{0}6.049\pm0.048$ \\
 -\phantom{0} & -                & $8.06\pm0.19$ & $7.1\pm0.1\phantom{0}$  & &  -$\phantom{0}$ & $9.54\pm0.16$ & $10.235\pm0.052$ \\
\end{tabular}
\end{ruledtabular}
\end{table*}
In other words, progressive improvements of mass resolution in reaction spectroscopy rendered access to detailed structures of $^{12}_\Lambda$C and $^{12}_\Lambda$B.
However, the reported excitation energies, which are summarized in Table~\ref{state_table}, show significant differences with one another, even though the global structure of spectra agree well.
Thus, precise reference data on the $^{12}_\Lambda$C level scheme determined by an independent method such as $\gamma$-ray spectroscopy are essential to all the reaction spectroscopy methods.

High precision $\gamma$-ray spectroscopy with a typical resolution of a few keV employing a dedicated germanium (Ge) detector array, Hyperball~\cite{tamura_2000_prl}, has been available since 1998 and revealed level schemes of various $p$-shell hypernuclei~\cite{hashimoto_2006_ppnp} below nucleon emission threshold.
The hypernuclear $\gamma$-ray data have been used to investigate spin-dependent parts of the $\Lambda$N interaction. 
In addition, precise level schemes determined from $\gamma$ rays have provided excellent reference data to the reaction spectroscopy experiments.

We report here the first precise measurement of the $^{12}_{\Lambda}$C level structure by $\gamma$-ray spectroscopy.
Preliminary results have been reported in Refs~\cite{ma_2007_epj,ma_2010_npa,hosomi_2013_npa}.

The KEK-PS E566 experiment was carried out at the K6 beam line of the 12\,GeV Proton Synchrotron (PS) in KEK.
For hypernuclear production, we used the $(\pi^{+}, K^{+})$ reaction at $1.05\,\text{GeV/$c$}$.
A total of $2 \times 10^{12}$ pions were irradiated on a $^{12}$C target of a $19.1\,\text{g/cm$^{2}$}$-thick polyethylene disk in one month beam time.
A typical beam intensity at the target was $3 \times 10^{6}$ particles per spill of 1.5\,s duration in every 4\,s.
Trajectories and momenta of beam pions and outgoing kaons were measured by the K6 beam spectrometer and Superconducting Kaon Spectrometer (SKS), respectively.
SKS had a large acceptance for detecting the outgoing kaons with laboratory scattering angles of $\theta_{\pi K}=0$--$20^{\circ}$.
More descriptions of the spectrometer system and analysis procedures for calculating missing mass are found in Refs.~\cite{fukuda_1995_nim, hotch_2001_prc}.


An array of Ge detectors, Hyperball2, was newly constructed and installed surrounding the target for $\gamma$-ray detection~\cite{ma_2007_epj}.
The total solid angle and the photo-peak efficiency of Hyperball2 were about $30\text{\%} \times 4\pi\,\text{sr}$ and 4\% for 1-MeV $\gamma$ rays, respectively. 
Energy calibration was performed over the range of 0.1--6.1\,MeV using a $^{152}$Eu source as well as $\gamma$ rays from surrounding materials activated by beam-induced reactions, such as $^{24}$Na(2754\,keV) from $^{27}\text{Al}(n,\alpha)$ and $^{16}$N(6129\,keV) from $^{16}\text{O}(n,p)$.
The systematic error in the energy calibration was estimated to be 0.3\,keV for the energy region below 3\,MeV and 0.6\,keV for the energy region above 3\,MeV. 
Performance of each Ge detector was continuously monitored using a $^{60}$Co source embedded in a plastic scintillation counter~\cite{tamura_2000_prl}.
The in-beam live time and the energy resolution were $(59.3\pm 0.3)$\% on average and 5.4\,keV (FWHM) at 1.33\,MeV, respectively.

In the $(\pi^{+}, K^{+})$ reaction at 1.05\,GeV/$c$, produced hypernuclei have recoil velocities of $\beta=0.028$--$0.038$, which lead to a typical stopping time of 2\,ps in the target medium. 
Therefore, $M1$ transitions with an energy lager than a few hundred keV are expected to have a broadened peak shape due to the Doppler-shift effect. 
We applied an event-by-event correction to $\gamma$-ray energy by using recoil momenta of $^{12}_{\Lambda}$C, reaction vertices, and positions of Ge detectors with hits.
It is noted that the Doppler-shift correction systematically gives a 1\% uncertainty on the measured $\gamma$-ray energy, where the dominant component is geometrical ambiguity ($\pm5$\,mm) in the positions of the Hyperball2 apparatus relative to the magnetic spectrometer system.
\begin{figure}
\includegraphics[width=8cm]{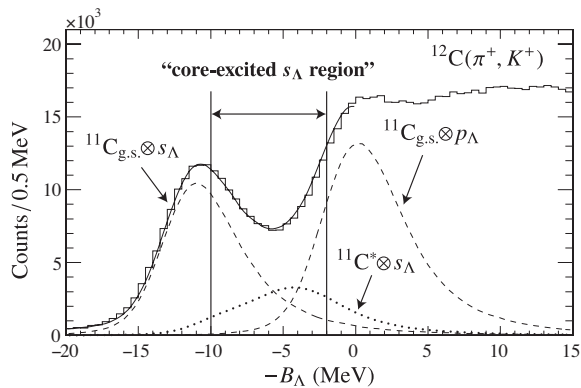}
\caption{\label{mass}$^{12}_{\Lambda}$C excitation spectrum obtained in the $^{12}$C$(\pi^{+}, K^{+})$ reaction with the missing mass resolution of 6\, MeV (FWHM). The dashed and the dotted curves show decomposition of $^{12}_{\Lambda}$C states based on a simulation (see text). The ``core-excited region" is defined as $-10<-B_{\Lambda}<-2$\,MeV for $\gamma$-ray analysis.}
\end{figure}


\begin{figure*}
\includegraphics[width=\textwidth]{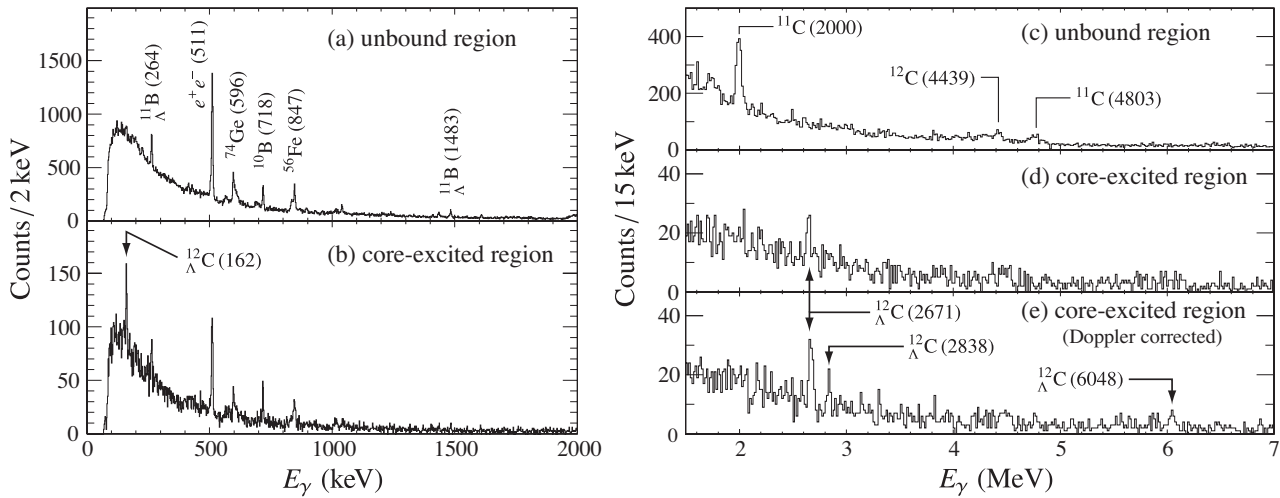}
\caption{\label{gamma}$\gamma$-ray energy spectra measured by Hyperball2 in coincidence with the $^{12}$C$(\pi^{+}, K^{+})$ reaction. Missing mass selections are applied to the unbound region ($-B_{\Lambda}>-2$\,MeV) for (a) and (c), and to the core-excited $s_{\Lambda}$ region ($-10<-B_{\Lambda}<-2$\,MeV) for (b), (d), and (e). The peak at 162\,keV in (b) is assigned as the direct $M1$ transition between the ground-state doublet ($2^{-}_{1} \to 1^{-}_{1}$). The spectrum (e) is obtained by applying an event-by-event Doppler-shift correction to the same data set of (d). Three peaks are observed in (e); two peaks at around 2.8\,MeV are attributed to the $M1(1^{-}_{2} \to 2^{-}_{1}, 1^{-}_{1})$ transitions, and the peak at 6\,MeV to the $M1(1^{-}_{3} \to 1^{-}_{1})$ transition.}
\end{figure*}

Figure~\ref{mass} shows the missing mass spectrum for $^{12}_{\Lambda}$C as a function of the $\Lambda$ binding energy $(B_{\Lambda})$.
As illustrated with the dashed and the dotted curves, $^{12}_{\Lambda}$C states are decomposed into three groups corresponding to different coupling of a $^{11}$C core to a $\Lambda$ hyperon orbit.
The dashed curve on the left side represents the $^{12}_{\Lambda}$C ground state.
The other curves are mixtures of several states, of which cross sections and excitation energies measured in KEK-E369~\cite{hotch_2001_prc} are referred.
We simulated a response function of a single state in our missing mass spectrum by taking account of the measured performance of the spectrometer system and the effect of target thickness.
As a result, the missing mass resolution was 6\,MeV (FWHM).
The absolute mass scale was adjusted so that the ground state has $B_{\Lambda}=10.76$\,MeV~\cite{davis_1992_npa}.
The region of $-10<-B_{\Lambda}<-2$\,MeV was set for a tight event selection of the core-excited $s_{\Lambda}$ states that are supposed to $\gamma$ decay.

Figure~\ref{gamma} shows $\gamma$-ray energy spectra. 
Figure~\ref{gamma}(a) and \ref{gamma}(c) are the spectra without the Doppler-shift correction when the unbound region ($-B_{\Lambda}>-2$\,MeV) of $^{12}_{\Lambda}$C is selected.
The $e^{+}e^{-}$ annihilation peak and several $\gamma$ rays from ordinary nuclei such as $^{10}$B, $^{11}$C, $^{12}$C, $^{56}$Fe, and $^{74}$Ge are present in the spectrum. 
Two peaks at 264\,keV and 1483\,keV are consistent with the known energies of transitions in $^{11}_{\Lambda}$B, which were previously measured with Hyperball in the $^{11}$B$(\pi^{+}, K^{+})$ reaction~\cite{miura_2005_npa}.
In the present experiment, $^{11}_{\Lambda}$B was formed through one proton emission from the unbound states of $^{12}_{\Lambda}$C. 
Figure~\ref{gamma}(b) and \ref{gamma}(d) are the spectra without the Doppler-shift correction for the ``core-excited $s_{\Lambda}$ region'', where peaks appear more prominently at 162\,keV and at 2671\,keV.
After the event-by-event Doppler-shift correction, the 2671-keV peak becomes narrower and then enhanced as shown in Fig.~\ref{gamma}(e).
Additionally, two peaks at 2838\,keV and 6048\,keV are observed with a statistical significance of about 3$\sigma$ in Fig.~\ref{gamma}(e).
The fact that the peaks are observed only in Fig.~\ref{gamma}(b) and in Fig.~\ref{gamma}(e) clearly demonstrates their origin in $^{12}_{\Lambda}$C.

The two peaks at 2671\,keV and 2838\,keV are assigned to transitions from the upper $1^{-}_{2}$ state to the ground-state doublet, by comparing their energies from the high-resolution $(\pi^{+}, K^{+})$ $^{12}_{\Lambda}$C spectrum with a thin target~\cite{hotch_2001_prc}.
In the weak-coupling picture, both transitions are of $M1$ character induced by the transition of the core nucleus itself, $^{11}\text{C}:M1(1/2^{-}\to3/2^{-})$.
It is consistent with the fact that the peaks are enhanced by the Doppler-shift correction.
As the results of fitting peaks in the Doppler-shift corrected spectrum, the $\gamma$-ray energies and the yields are $2671\pm3(\text{stat})\pm3(\text{syst})$\,keV and $56\pm10$ counts for the lower peak, and  $2838\pm4\pm3$\,keV and $24\pm7$ counts for the upper peak, respectively.
Since the transition ratio of $(1^{-}_{2}\to1^{-}_{1})/(1^{-}_{2}\to2^{-}_{1})$ is calculated to be 0.33 in the weak coupling limit, the observed yield ratio $N(2838)/N(2671)=0.43\pm0.15$ strongly supports a spin $1^{-}$ assignment to the ground state.
This is further evidenced by a $\gamma$-ray yield for the direct $M1$ transition between the ground-state doublet as discussed below.

The energy difference ($167\pm5$\,keV) between the 2671\,keV and 2838\,keV transitions overlaps with the 162-keV peak energy.
Thus, the 162-keV $\gamma$ ray is assigned as the $M1$ transition between the ground-state doublet $(2^{-}_{1}\to1^{-}_{1})$, where the essential process is the spin flip of a $\Lambda$ hyperon.
A simple gaussian fitting gives the transition energy of $161.5\pm0.3\pm0.3$\,keV and the yield of $172\pm25$ counts.
The $2^{-}_{1}$ state is excited by the spin-flip interaction in the $(\pi^{+}, K^{+})$ reaction with a small cross section relative to the spin-non-flip $1^{-}_{1}$ state (about 10\% at $\theta_{\pi K}=15^{\circ}$) according to theoretical prediction~\cite{itonaga_1994_prc}. 
The observed 162-keV yield is consistent with the expected yield for the $(2^{-}_{1}\to 1^{-}_{1})$ transition via the direct spin-flip production of the $2^{-}_{1}$ state and the cascade feedings from the upper excited states.

The 6-MeV peak observed in the Doppler-shift corrected spectrum is attributed to the $M1(1^{-}_{3}\to1^{-}_{1})$ transition.
The $\gamma$-ray energy and yield of $6048\pm8\pm7$\,keV and $21\pm7$ counts, respectively, are obtained by fitting with a simulated response function.
The observed peak width agrees with a fully Doppler broadened width.
Although it is difficult to identify the population of the $1^{-}_{3}$ state in our missing mass spectrum, the $1^{-}_{3}$ state of $^{12}_{\Lambda}$C ($^{11}_{\Lambda}$B) has been confirmed at $\sim$6\,MeV excitation energy in other experiments with much better missing mass resolutions (see Table~\ref{state_table}).
The $M1(1^{-}_{3}\to1^{-}_{1})$ yield is calculated to be $24\pm6$ counts based on the $1^{-}_{3}$ state cross section measured in KEK-E369, theoretical $\gamma$-ray branching ratios~\cite{itonaga_1994_ptps}, and detector efficiencies in the present experiment.
The obtained yield of the 6-MeV $\gamma$ ray supports the assignment of $M1(1^{-}_{3}\to1^{-}_{1})$ to this transition.
 
The excitation energies of the $^{12}_{\Lambda}$C states were determined by applying a nuclear recoil correction to the observed $\gamma$-ray energies.
They are tabulated in Table~\ref{state_table} with comparable results from other experiments.
For the 2.8-MeV excitation energy, we took an energy sum of the cascading transitions $(1^{-}_{2}\to2^{-}_{1}\to1^{-}_{1})$ because of a poor statistics for the ($1^{-}_{2}\to1^{-}_{1}$) transition.

\begin{figure}
\includegraphics[width=8cm]{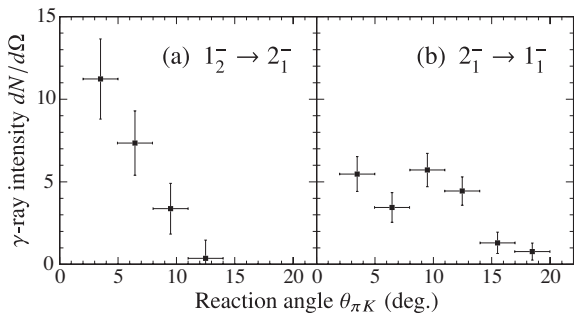}
\caption{\label{angle} Reaction angle ($\theta_{\pi K}$) dependence of the $^{12}_{\Lambda}$C $\gamma$-ray intensities measured in the $(\pi^{+}, K^{+})$ reaction at 1.05\,GeV/$c$. The vertical axis shows relative intensities of $\gamma$ rays that are calculated from the measured $\gamma$-ray counts with corrections (see text).}
\end{figure}

Figure~\ref{angle} shows relative intensities of $\gamma$-ray transitions plotted against the $(\pi^{+}, K^{+})$ reaction angle with an interval of $3^{\circ}$ in the laboratory frame.
The relative intensities were calculated from the measured $\gamma$-ray counts with a correction for the spectrometer acceptance and with a normalization for the $\gamma$-ray efficiency at 162\,keV.
We compared the obtained angular distributions with the calculations~\cite{bando_1989_prc} in terms of the spin-flip and the spin-non-flip production components.
The $(1^{-}_{2}\to1^{-}_{1})$ intensity curve is well understood based on the calculation for the $1^{-}_{2}$ spin-non-flip state.
On the other hand, the $(2^{-}_{1}\to1^{-}_{1})$ intensities show an enhancement at around $\theta_{\pi K}=10^{\circ}$, at which the calculation for the $2^{-}$ spin-flip state also has the maximum amplitude.
This enhancement indicates a contribution of the direct $2^{-}_{1}$ population in addition to the feeding process from the upper excited states.


With an assumption that the $2^{-}_{1}$ state is populated by two processes, namely the direct production and the feeding from the $1^{-}_{2}$ state, the production ratio of $2^{-}_{1}/1^{-}_{1}$ was estimated to be $(8\pm3)$\% at $\theta_{\pi K}=2^{\circ}$--$20^{\circ}$ from the yields of the cascading $(1^{-}_{2}\to2^{-}_{1}\to1^{-}_{1})$ transitions, the cross sections of ($1^{-}_{1}$, $1^{-}_{2}$) in KEK-E369, and an estimated weak decay branching ratio of the $2^{-}_{1}$ state $(\Gamma_{M1}:\Gamma_{\text{weak}}=0.32:0.68)$~\footnote{For the estimation of the weak decay branching ratio, the calculated $M1$ transition width~\cite{dalitz_1978_aop} and the measured weak decay width~\cite{kameoka_2005_npa} for the ground state were used. The magnetic moments are referred in Ref~\cite{toi_npa} for a $^{11}$C nucleus and in Ref~\cite{pdg} for a $\Lambda$ hyperon. }.
More elaborate analysis gave the $2^{-}_{1}/1^{-}_{1}$ ratio of $(5\pm3)$\% with a consideration of feeding from possible excited states.
The theoretical distributions predict 6\% for the $2^{-}_{1}/1^{-}_{1}$ ratio in the region of $\theta_{\pi K}=2^{\circ}$--$20^{\circ}$.
The calculations agree with the present result.



The measured energy spacing of the $(2^{-}_{1},1^{-}_{1})$ doublet implies the fact that weak decay competes with the $M1$ transition for the $2^{-}_{1}$ state. 
This poses a question in weak decay measurements of $^{12}_{\Lambda}$C~\cite{kim_2003_prc, okada_2004_plb} in which contamination from the $2^{-}_{1}$ weak decay was not taken into account.


The energy spacing itself also offers new data for the $\Lambda N$ interaction.
Low-lying levels of $p$-shell hypernuclei have been studied with a phenomenological approach, where strengths of the spin-dependent $\Lambda N$ interactions are parametrized in four terms~\cite{gal_1971_aop, millener_1985_prc}.
A recent shell model calculation~\cite{millener_2012_npa} predicts a comparable doublet spacing of 153\,keV by using the already known $\Lambda$N interaction parameters and an additional term for the $\Lambda$-$\Sigma$ coupling effect.


In summary, the KEK-PS E566 experiment successfully identified four $\gamma$-ray transitions from $^{12}_{\Lambda}$C produced by the $^{12}$C$(\pi^{+},K^{+})$ reaction and deduced the low-lying level scheme of $^{12}_{\Lambda}$C with precise excitation energies.
The obtained $(2^{-}_{1},1^{-}_{1})$ doublet spacing is consistent with other $p$-shell data in terms of strengths of the spin-dependent $\Lambda$N interactions. 
In spectroscopic studies of $\Lambda$ hypernuclei employing various reaction methods, the $^{12}_{\Lambda}$C ($^{12}_{\Lambda}$B) excitation spectra have played a particularly important role in evaluation and verification of the spectrometer performance as well as in investigation of reaction mechanism and weak decay properties of $\Lambda$ hypernuclei.
The $\gamma$-ray data obtained in the present work have an impact on the past and future experiments in hypernuclear spectroscopy by providing a solid reference to the benchmark spectra.
\begin{acknowledgments}
We acknowledge experimental support from the KEK Proton Synchrotron (KEK-PS) staff.
This work is partially supported by Grant-in-Aid from JSPS and MEXT (Nos. 17070001, 23244043, and 24105003).
\end{acknowledgments}

\bibliography{e566_prl}

\begin{thebibliography}{31}%
\makeatletter
\providecommand \@ifxundefined [1]{%
 \@ifx{#1\undefined}
}%
\providecommand \@ifnum [1]{%
 \ifnum #1\expandafter \@firstoftwo
 \else \expandafter \@secondoftwo
 \fi
}%
\providecommand \@ifx [1]{%
 \ifx #1\expandafter \@firstoftwo
 \else \expandafter \@secondoftwo
 \fi
}%
\providecommand \natexlab [1]{#1}%
\providecommand \enquote  [1]{``#1''}%
\providecommand \bibnamefont  [1]{#1}%
\providecommand \bibfnamefont [1]{#1}%
\providecommand \citenamefont [1]{#1}%
\providecommand \href@noop [0]{\@secondoftwo}%
\providecommand \href [0]{\begingroup \@sanitize@url \@href}%
\providecommand \@href[1]{\@@startlink{#1}\@@href}%
\providecommand \@@href[1]{\endgroup#1\@@endlink}%
\providecommand \@sanitize@url [0]{\catcode `\\12\catcode `\$12\catcode
  `\&12\catcode `\#12\catcode `\^12\catcode `\_12\catcode `\%12\relax}%
\providecommand \@@startlink[1]{}%
\providecommand \@@endlink[0]{}%
\providecommand \url  [0]{\begingroup\@sanitize@url \@url }%
\providecommand \@url [1]{\endgroup\@href {#1}{\urlprefix }}%
\providecommand \urlprefix  [0]{URL }%
\providecommand \Eprint [0]{\href }%
\providecommand \doibase [0]{http://dx.doi.org/}%
\providecommand \selectlanguage [0]{\@gobble}%
\providecommand \bibinfo  [0]{\@secondoftwo}%
\providecommand \bibfield  [0]{\@secondoftwo}%
\providecommand \translation [1]{[#1]}%
\providecommand \BibitemOpen [0]{}%
\providecommand \bibitemStop [0]{}%
\providecommand \bibitemNoStop [0]{.\EOS\space}%
\providecommand \EOS [0]{\spacefactor3000\relax}%
\providecommand \BibitemShut  [1]{\csname bibitem#1\endcsname}%
\let\auto@bib@innerbib\@empty
\bibitem [{\citenamefont {{M.\ A.\ Faessler
  $et~al.$}}(1973)}]{faessler_1973_pl}%
  \BibitemOpen
  \bibfield  {author} {\bibinfo {author} {\bibnamefont {{M.\ A.\ Faessler
  $et~al.$}}},\ }\href@noop {} {\bibfield  {journal} {\bibinfo  {journal}
  {Phys.\ Lett. B}\ }\textbf {\bibinfo {volume} {46}},\ \bibinfo {pages} {468}
  (\bibinfo {year} {1973})}\BibitemShut {NoStop}%
\bibitem [{\citenamefont {{W.\ Br\"{u}ckner
  $et~al.$}}(1975)}]{bruckner_1975_plb}%
  \BibitemOpen
  \bibfield  {author} {\bibinfo {author} {\bibnamefont {{W.\ Br\"{u}ckner
  $et~al.$}}},\ }\href@noop {} {\bibfield  {journal} {\bibinfo  {journal}
  {Phys.\ Lett.\ B}\ }\textbf {\bibinfo {volume} {55}},\ \bibinfo {pages} {107}
  (\bibinfo {year} {1975})}\BibitemShut {NoStop}%
\bibitem [{\citenamefont {{C.\ Milner $et~al.$}}(1985)}]{milner_1985_prl}%
  \BibitemOpen
  \bibfield  {author} {\bibinfo {author} {\bibnamefont {{C.\ Milner
  $et~al.$}}},\ }\href@noop {} {\bibfield  {journal} {\bibinfo  {journal}
  {Phys.\ Rev.\ Lett.}\ }\textbf {\bibinfo {volume} {54}},\ \bibinfo {pages}
  {1237} (\bibinfo {year} {1985})}\BibitemShut {NoStop}%
\bibitem [{\citenamefont {{M.\ Akei $et~al.$}}(1991)}]{akei_1991_npa}%
  \BibitemOpen
  \bibfield  {author} {\bibinfo {author} {\bibnamefont {{M.\ Akei $et~al.$}}},\
  }\href@noop {} {\bibfield  {journal} {\bibinfo  {journal} {Nucl.\ Phys.\ A}\
  }\textbf {\bibinfo {volume} {534}},\ \bibinfo {pages} {478} (\bibinfo {year}
  {1991})}\BibitemShut {NoStop}%
\bibitem [{\citenamefont {{T.\ Miyoshi $et~al.$}}(2003)}]{miyoshi_2003_prl}%
  \BibitemOpen
  \bibfield  {author} {\bibinfo {author} {\bibnamefont {{T.\ Miyoshi
  $et~al.$}}},\ }\href@noop {} {\bibfield  {journal} {\bibinfo  {journal}
  {Phys.\ Rev.\ Lett.}\ }\textbf {\bibinfo {volume} {90}},\ \bibinfo {pages}
  {232502} (\bibinfo {year} {2003})}\BibitemShut {NoStop}%
\bibitem [{\citenamefont {{M.\ Agnello $et~al.$}}(2005)}]{agnello_2005_plb}%
  \BibitemOpen
  \bibfield  {author} {\bibinfo {author} {\bibnamefont {{M.\ Agnello
  $et~al.$}}},\ }\href@noop {} {\bibfield  {journal} {\bibinfo  {journal}
  {Phys.\ Lett.\ B}\ }\textbf {\bibinfo {volume} {622}},\ \bibinfo {pages} {35}
  (\bibinfo {year} {2005})}\BibitemShut {NoStop}%
\bibitem [{\citenamefont {{T.\ Hasegawa $et~al.$}}(1995)}]{hasegawa_1995_prl}%
  \BibitemOpen
  \bibfield  {author} {\bibinfo {author} {\bibnamefont {{T.\ Hasegawa
  $et~al.$}}},\ }\href@noop {} {\bibfield  {journal} {\bibinfo  {journal}
  {Phys.\ Rev.\ Lett.}\ }\textbf {\bibinfo {volume} {74}},\ \bibinfo {pages}
  {224} (\bibinfo {year} {1995})}\BibitemShut {NoStop}%
\bibitem [{\citenamefont {{H.\ Hotchi $et~al.$}}(2001)}]{hotch_2001_prc}%
  \BibitemOpen
  \bibfield  {author} {\bibinfo {author} {\bibnamefont {{H.\ Hotchi
  $et~al.$}}},\ }\href@noop {} {\bibfield  {journal} {\bibinfo  {journal}
  {Phys.\ Rev.\ C}\ }\textbf {\bibinfo {volume} {64}},\ \bibinfo {pages}
  {044302} (\bibinfo {year} {2001})}\BibitemShut {NoStop}%
\bibitem [{\citenamefont {{M.\ Iodice $et~al.$}}(2007)}]{iodice_2007_prl}%
  \BibitemOpen
  \bibfield  {author} {\bibinfo {author} {\bibnamefont {{M.\ Iodice
  $et~al.$}}},\ }\href@noop {} {\bibfield  {journal} {\bibinfo  {journal}
  {Phys.\ Rev.\ Lett.}\ }\textbf {\bibinfo {volume} {99}},\ \bibinfo {pages}
  {052501} (\bibinfo {year} {2007})}\BibitemShut {NoStop}%
\bibitem [{\citenamefont {{L.\ Tang $et~al.$}}(2014)}]{tang_2014_prc}%
  \BibitemOpen
  \bibfield  {author} {\bibinfo {author} {\bibnamefont {{L.\ Tang $et~al.$}}},\
  }\href@noop {} {\bibfield  {journal} {\bibinfo  {journal} {Phys.\ Rev.\ C}\
  }\textbf {\bibinfo {volume} {90}},\ \bibinfo {pages} {034320} (\bibinfo
  {year} {2014})}\BibitemShut {NoStop}%
\bibitem [{\citenamefont {{H.\ Tamura $et~al.$}}(2000)}]{tamura_2000_prl}%
  \BibitemOpen
  \bibfield  {author} {\bibinfo {author} {\bibnamefont {{H.\ Tamura
  $et~al.$}}},\ }\href@noop {} {\bibfield  {journal} {\bibinfo  {journal}
  {Phys.\ Rev.\ Lett.}\ }\textbf {\bibinfo {volume} {84}},\ \bibinfo {pages}
  {5963} (\bibinfo {year} {2000})}\BibitemShut {NoStop}%
\bibitem [{\citenamefont {{O.\ Hashimoto and H.\
  Tamura}}(2006)}]{hashimoto_2006_ppnp}%
  \BibitemOpen
  \bibfield  {author} {\bibinfo {author} {\bibnamefont {{O.\ Hashimoto and H.\
  Tamura}}},\ }\href@noop {} {\bibfield  {journal} {\bibinfo  {journal} {Prog.\
  Part.\ Nucl.\ Phys.}\ }\textbf {\bibinfo {volume} {57}},\ \bibinfo {pages}
  {564} (\bibinfo {year} {2006})}\BibitemShut {NoStop}%
\bibitem [{\citenamefont {{Y.\ Ma $et~al.$}}(2007)}]{ma_2007_epj}%
  \BibitemOpen
  \bibfield  {author} {\bibinfo {author} {\bibnamefont {{Y.\ Ma $et~al.$}}},\
  }\href@noop {} {\bibfield  {journal} {\bibinfo  {journal} {Eur.\ Phys.\ J.\
  A}\ }\textbf {\bibinfo {volume} {33}},\ \bibinfo {pages} {243} (\bibinfo
  {year} {2007})}\BibitemShut {NoStop}%
\bibitem [{\citenamefont {{Y.\ Ma $et~al.$}}(2010)}]{ma_2010_npa}%
  \BibitemOpen
  \bibfield  {author} {\bibinfo {author} {\bibnamefont {{Y.\ Ma $et~al.$}}},\
  }\href@noop {} {\bibfield  {journal} {\bibinfo  {journal} {Nucl.\ Phys.\ A}\
  }\textbf {\bibinfo {volume} {835}},\ \bibinfo {pages} {422} (\bibinfo {year}
  {2010})}\BibitemShut {NoStop}%
\bibitem [{\citenamefont {{K.\ Hosomi $et~al.$}}(2013)}]{hosomi_2013_npa}%
  \BibitemOpen
  \bibfield  {author} {\bibinfo {author} {\bibnamefont {{K.\ Hosomi
  $et~al.$}}},\ }\href@noop {} {\bibfield  {journal} {\bibinfo  {journal}
  {Nucl.\ Phys.\ A}\ }\textbf {\bibinfo {volume} {914}},\ \bibinfo {pages}
  {184} (\bibinfo {year} {2013})}\BibitemShut {NoStop}%
\bibitem [{\citenamefont {{F.\ Ajzenberg-Selove}}(1990)}]{toi_npa}%
  \BibitemOpen
  \bibfield  {author} {\bibinfo {author} {\bibnamefont {{F.\
  Ajzenberg-Selove}}},\ }\href@noop {} {\bibfield  {journal} {\bibinfo
  {journal} {Nucl.\ Phys.\ A}\ }\textbf {\bibinfo {volume} {506}},\ \bibinfo
  {pages} {1} (\bibinfo {year} {1990})}\BibitemShut {NoStop}%
\bibitem [{\citenamefont {{T.\ Fukuda $et~al.$}}(1995)}]{fukuda_1995_nim}%
  \BibitemOpen
  \bibfield  {author} {\bibinfo {author} {\bibnamefont {{T.\ Fukuda
  $et~al.$}}},\ }\href@noop {} {\bibfield  {journal} {\bibinfo  {journal}
  {Nucl.\ Instr.\ Meth.\ A}\ }\textbf {\bibinfo {volume} {361}},\ \bibinfo
  {pages} {485} (\bibinfo {year} {1995})}\BibitemShut {NoStop}%
\bibitem [{\citenamefont {{D.\ H.\ Davis}}(1992)}]{davis_1992_npa}%
  \BibitemOpen
  \bibfield  {author} {\bibinfo {author} {\bibnamefont {{D.\ H.\ Davis}}},\
  }\href@noop {} {\bibfield  {journal} {\bibinfo  {journal} {Nucl.\ Phys.\ A}\
  }\textbf {\bibinfo {volume} {547}},\ \bibinfo {pages} {369} (\bibinfo {year}
  {1992})}\BibitemShut {NoStop}%
\bibitem [{\citenamefont {{Y.\ Miura $et~al.$}}(2005)}]{miura_2005_npa}%
  \BibitemOpen
  \bibfield  {author} {\bibinfo {author} {\bibnamefont {{Y.\ Miura
  $et~al.$}}},\ }\href@noop {} {\bibfield  {journal} {\bibinfo  {journal}
  {Nucl.\ Phys.\ A}\ }\textbf {\bibinfo {volume} {754}},\ \bibinfo {pages} {75}
  (\bibinfo {year} {2005})}\BibitemShut {NoStop}%
\bibitem [{\citenamefont {{K.\ Itonaga, T.\ Motoba, O.\ Richter and M.\
  Sotona}}(1994)}]{itonaga_1994_prc}%
  \BibitemOpen
  \bibfield  {author} {\bibinfo {author} {\bibnamefont {{K.\ Itonaga, T.\
  Motoba, O.\ Richter and M.\ Sotona}}},\ }\href@noop {} {\bibfield  {journal}
  {\bibinfo  {journal} {Phys.\ Rev.\ C}\ }\textbf {\bibinfo {volume} {49}},\
  \bibinfo {pages} {1045} (\bibinfo {year} {1994})}\BibitemShut {NoStop}%
\bibitem [{\citenamefont {{K.\ Itonaga, T.\ Motoba and M.\
  Sotona}}(1994)}]{itonaga_1994_ptps}%
  \BibitemOpen
  \bibfield  {author} {\bibinfo {author} {\bibnamefont {{K.\ Itonaga, T.\
  Motoba and M.\ Sotona}}},\ }\href@noop {} {\bibfield  {journal} {\bibinfo
  {journal} {Prog.\ Theor.\ Phys.\ Suppl.}\ }\textbf {\bibinfo {volume}
  {117}},\ \bibinfo {pages} {17} (\bibinfo {year} {1994})}\BibitemShut
  {NoStop}%
\bibitem [{\citenamefont {{H.\ Band{\={o}}, T.\ Motoba, M.\ Sotona and J.
  \v{Z}ofka}}(1989)}]{bando_1989_prc}%
  \BibitemOpen
  \bibfield  {author} {\bibinfo {author} {\bibnamefont {{H.\ Band{\={o}}, T.\
  Motoba, M.\ Sotona and J. \v{Z}ofka}}},\ }\href@noop {} {\bibfield  {journal}
  {\bibinfo  {journal} {Phys.\ Rev.\ C}\ }\textbf {\bibinfo {volume} {39}},\
  \bibinfo {pages} {587} (\bibinfo {year} {1989})}\BibitemShut {NoStop}%
\bibitem [{Note1()}]{Note1}%
  \BibitemOpen
  \bibinfo {note} {For the estimation of the weak decay branching ratio, the
  calculated $M1$ transition width~\cite {dalitz_1978_aop} and the measured
  weak decay width~\cite {kameoka_2005_npa} for the ground state were used. The
  magnetic moments are referred in Ref~\cite {toi_npa} for a $^{11}$C nucleus
  and in Ref~\cite {pdg} for a $\Lambda $ hyperon.}\BibitemShut {Stop}%
\bibitem [{\citenamefont {{J.\ H.\ Kim $et~al.$}}(2003)}]{kim_2003_prc}%
  \BibitemOpen
  \bibfield  {author} {\bibinfo {author} {\bibnamefont {{J.\ H.\ Kim
  $et~al.$}}},\ }\href@noop {} {\bibfield  {journal} {\bibinfo  {journal}
  {Phys.\ Rev.\ C}\ }\textbf {\bibinfo {volume} {68}},\ \bibinfo {pages}
  {065201} (\bibinfo {year} {2003})}\BibitemShut {NoStop}%
\bibitem [{\citenamefont {{S.\ Okada $et~al.$}}(2004)}]{okada_2004_plb}%
  \BibitemOpen
  \bibfield  {author} {\bibinfo {author} {\bibnamefont {{S.\ Okada
  $et~al.$}}},\ }\href@noop {} {\bibfield  {journal} {\bibinfo  {journal}
  {Phys.\ Lett.\ B}\ }\textbf {\bibinfo {volume} {597}},\ \bibinfo {pages}
  {249} (\bibinfo {year} {2004})}\BibitemShut {NoStop}%
\bibitem [{\citenamefont {{A.\ Gal, J.\ M.\ Soper and R.\ H.\
  Dalitz}}(1971)}]{gal_1971_aop}%
  \BibitemOpen
  \bibfield  {author} {\bibinfo {author} {\bibnamefont {{A.\ Gal, J.\ M.\ Soper
  and R.\ H.\ Dalitz}}},\ }\href@noop {} {\bibfield  {journal} {\bibinfo
  {journal} {Ann.\ Phys.}\ }\textbf {\bibinfo {volume} {63}},\ \bibinfo {pages}
  {53} (\bibinfo {year} {1971})}\BibitemShut {NoStop}%
\bibitem [{\citenamefont {{D.\ J.\ Millener, A.\ Gal, C.\ B.\ Dover and R.\ H.
  Dalitz}}(1985)}]{millener_1985_prc}%
  \BibitemOpen
  \bibfield  {author} {\bibinfo {author} {\bibnamefont {{D.\ J.\ Millener, A.\
  Gal, C.\ B.\ Dover and R.\ H. Dalitz}}},\ }\href@noop {} {\bibfield
  {journal} {\bibinfo  {journal} {Phys.\ Rev.\ C}\ }\textbf {\bibinfo {volume}
  {31}},\ \bibinfo {pages} {499} (\bibinfo {year} {1985})}\BibitemShut
  {NoStop}%
\bibitem [{\citenamefont {{D.\ J.\ Millener}}(2012)}]{millener_2012_npa}%
  \BibitemOpen
  \bibfield  {author} {\bibinfo {author} {\bibnamefont {{D.\ J.\ Millener}}},\
  }\href@noop {} {\bibfield  {journal} {\bibinfo  {journal} {Nucl.\ Phys.\ A}\
  }\textbf {\bibinfo {volume} {881}},\ \bibinfo {pages} {298} (\bibinfo {year}
  {2012})}\BibitemShut {NoStop}%
\bibitem [{\citenamefont {{R.\ H.\ Dalitz and A.\
  Gal}}(1978)}]{dalitz_1978_aop}%
  \BibitemOpen
  \bibfield  {author} {\bibinfo {author} {\bibnamefont {{R.\ H.\ Dalitz and A.\
  Gal}}},\ }\href@noop {} {\bibfield  {journal} {\bibinfo  {journal} {Ann.\
  Phys.}\ }\textbf {\bibinfo {volume} {116}},\ \bibinfo {pages} {167} (\bibinfo
  {year} {1978})}\BibitemShut {NoStop}%
\bibitem [{\citenamefont {{S.\ Kameoka $et~al.$}}(2005)}]{kameoka_2005_npa}%
  \BibitemOpen
  \bibfield  {author} {\bibinfo {author} {\bibnamefont {{S.\ Kameoka
  $et~al.$}}},\ }\href@noop {} {\bibfield  {journal} {\bibinfo  {journal}
  {Nucl.\ Phys.\ A}\ }\textbf {\bibinfo {volume} {754}},\ \bibinfo {pages}
  {173} (\bibinfo {year} {2005})}\BibitemShut {NoStop}%
\bibitem [{\citenamefont {{K.\ A.\ Olive $et~al.$}}(2014)}]{pdg}%
  \BibitemOpen
  \bibfield  {author} {\bibinfo {author} {\bibnamefont {{K.\ A.\ Olive
  $et~al.$}}},\ }\href@noop {} {\bibfield  {journal} {\bibinfo  {journal}
  {Chin.\ Phys.\ C}\ }\textbf {\bibinfo {volume} {38}},\ \bibinfo {pages}
  {090001} (\bibinfo {year} {2014})}\BibitemShut {NoStop}%
\end{thebibliography}%

\end{document}